\documentclass[lettersize,journal]{IEEEtran}

\usepackage{algorithm}
\usepackage{algpseudocode}
\usepackage{amsmath,amsfonts}
\usepackage{array}
\usepackage[caption=false,font=normalsize,labelfont=sf,textfont=sf]{subfig}
\usepackage{textcomp}
\usepackage{stfloats}
\usepackage{url}
\usepackage{verbatim}
\usepackage{graphicx}
\usepackage{cite}

\usepackage{booktabs}
\usepackage{makecell}
\usepackage{multirow}
\usepackage{multicol}
\usepackage{tabularx}
\usepackage[dvipsnames]{xcolor}
\usepackage{caption}
\usepackage{shellesc}
\usepackage{tablefootnote} 
\usepackage{nomencl}
\makenomenclature

\usepackage{hyperref}
\hypersetup{hidelinks}

\usepackage[acronym]{glossaries}

\newacronym{ss}{speaker separation}{speaker separation}
\newacronym{se}{speech enhancement}{speech enhancement}
\newacronym{tse}{TSE}{target speaker extraction}
\newacronym{sex}{SEx}{speaker extraction}

\newacronym{ao}{AO}{audio-only}
\newacronym{av}{AV}{audiovisual}
\newacronym{avse}{AVSE}{audiovisual speech enhancement}
\newacronym{avsr}{AVSR}{audiovisual speech recognition}
\newacronym{avss}{AVSS}{audiovisual speaker separation}
\newacronym{avtse}{AVTSE}{audiovisual target speaker extraction}

\newacronym{istft}{iSTFT}{inverse short-time Fourier transform}
\newacronym{stft}{STFT}{short-time Fourier transform}
\newacronym{tf}{T-F}{time-frequency}
\newacronym{rir}{RIR}{room impulse responses}
\newacronym{mcmfwf}{MCMFWF}{multi-channel multi-frame Wiener filter}
\newacronym{scmfwf}{SCMFWF}{single-channel multi-frame Wiener filter}
\newacronym{asr}{ASR}{automatic speech recognition}
\newacronym{ri}{RI}{real and imaginary}

\newacronym{ffn}{FFN}{feed-forward networks}
\newacronym{deconv1d}{Deconv1D}{one-dimensional deconvolution}
\newacronym{mhsa}{MHSA}{multi-head self-attention}
\newacronym{gmhsa}{GMHSA}{global multi-head self-attention}
\newacronym{ln}{LN}{layer normalization}
\newacronym{bn}{BN}{batch normalization}
\newacronym{relu}{ReLU}{rectified linear unit}
\newacronym{prelu}{PReLU}{parametric rectified linear unit}
\newacronym{lnorm}{LN}{layer normalization}
\newacronym{glnorm}{gLN}{global layer normalization}
\newacronym{gnorm}{GN}{group normalization}
\newacronym{conv2d}{Conv2D}{2D convolutional layer}
\newacronym{conv1d}{Conv1D}{1D convolutional layer}
\newacronym{dnn}{DNN}{deep neural network}
\newacronym{cnn}{CNN}{convolutional neural network}
\newacronym{rnn}{RNN}{recurrent neural network}
\newacronym{lstm}{LSTM}{long short-term memory}
\newacronym{blstm}{BLSTM}{bidirectional long short-term memory}
\newacronym{gru}{GRU}{gated recurrent unit}
\newacronym{sgd}{SGD}{stochastic gradient descent}
\newacronym{adam}{Adam}{adaptive moment estimation}
\newacronym{bptt}{BPTT}{backpropagation through time}
\newacronym{nlp}{NLP}{natural language processing}
\newacronym{cv}{CV}{computer vision}
\newacronym{dm}{DM}{dynamic mixing}
\newacronym{pe}{PE}{positional encoding}
\newacronym{rpe}{RPE}{relative positional encoding}
\newacronym{rcpe}{RCPE}{random-chunk positional encoding}
\newacronym{tconvffn}{T-ConvFFN}{time-convolutional feed-forward network}
\newacronym{tconv}{T-Conv}{time-convolutional}
\newacronym{tgcon1d}{T-GConv1D}{grouped 1d convolution}
\newacronym{vtcn}{V-TCN}{visual temporal convolutional network}

\newacronym{silu}{SiLU}{sigmoid-weighted linear unit}
\newacronym{pit}{PIT}{permutation invariant training}
\newacronym{vbt}{VBT}{visual based training}

\newacronym{nb-mhsa}{NB-MHSA}{narrow-band multi-head self-attention}

\newacronym{pesq}{PESQ}{perceptual evaluation of speech quality}
\newacronym{nb-pesq}{PESQ}{narrow-band perceptual evaluation of speech quality}
\newacronym{snr}{SNR}{signal-to-noise ratio}
\newacronym{tir}{TIR}{target-to-interferer ratio}
\newacronym{sdr}{SDR}{signal-to-distortion ratio}
\newacronym{sdri}{SDRi}{signal-to-distortion ratio improvement}
\newacronym{si-sdr}{SI-SDR}{scale-invariant signal-to-distortion ratio}
\newacronym{si-snri}{SI-SNRi}{scale-invariant signal-to-noise ratio improvement}
\newacronym{si-sdrse}{SI-SDR(SE)}{scale-invariant signal-to-distortion ratio (scaled estimated)}
\newacronym{si-sdri}{SI-SDRi}{scale-invariant signal-to-distortion ratio improvement}
\newacronym{stoi}{STOI}{short-time objective intelligibility} 
\newacronym{estoi}{eSTOI}{extended short-time objective intelligibility} 
\newacronym{mc}{MC}{mixture constraint loss}
\newacronym{wer}{WER}{word error rate}

\newacronym{fps}{FPS}{frames per second}

\newacronym{flop}{FLOP}{floating point operation}
\newacronym{gflop}{GFLOP}{Giga floating point operation}

\newacronym{lrs}{LRS}{lip reading sentences}

\usepackage{gensymb}

\usepackage{pifont}
\newcommand{\nommodule}[1]{%
\renewcommand{\nomentryend}{\hspace*{\fill}#1}}

\makenomenclature

\begin{document}

\title{AV-CrossNet: an Audiovisual Complex Spectral Mapping Network for Speech Separation By Leveraging Narrow- and Cross-Band Modeling}

\author{
Vahid Ahmadi Kalkhorani, Cheng Yu, Anurag Kumar, Ke Tan, Buye Xu,
DeLiang Wang, ~\IEEEmembership{Fellow,~IEEE}
\thanks{
Vahid Ahmadi Kalkhorani and Cheng Yu are with the Department of Computer Science and Engineering, Ohio State University, Columbus, OH 43210 USA (e-mail: \href{ahmadikalkhorani.1@osu.edu}{ahmadikalkhorani.1@osu.edu}; \href{yu.3500@osu.edu}{yu.3500@osu.edu}).
Anurag Kumar, Ke Tan, and Buye Xu are with Meta Reality Labs, Redmond, WA
20004 USA (e-mail: \href{anuragkr90@meta.com}{anuragkr90@meta.com}; \href{tanke1116@meta.com}{tanke1116@meta.com}; \href{xub@meta.com}{xub@meta.com}).
DeLiang Wang is with the Department of Computer Science and Engineering and the Center for Cognitive and Brain Sciences, Ohio State University, Columbus, OH 43210 USA (e-mail: \href{dwang@cse.ohio-state.edu}{dwang@cse.ohio-state.edu}).

}
}



\maketitle

\begin{abstract}
     Adding visual cues to audio-based speech separation can improve separation performance. 
 This paper introduces AV-CrossNet, an \gls{av} system for speech enhancement, target speaker extraction, and multi-talker speaker separation. AV-CrossNet is extended from the CrossNet architecture, which is a recently proposed network that performs complex spectral mapping for speech separation by leveraging global attention and positional encoding. To effectively utilize visual cues, the proposed system incorporates pre-extracted visual embeddings and employs a visual encoder comprising temporal convolutional layers. Audio and visual features are fused in an early fusion layer before feeding to AV-CrossNet blocks. We evaluate AV-CrossNet on multiple datasets, including LRS, VoxCeleb, and COG-MHEAR challenge. Evaluation results demonstrate that AV-CrossNet advances the state-of-the-art performance in all audiovisual tasks, even on untrained and mismatched datasets.
\end{abstract}

\begin{IEEEkeywords}
    Audiovisual speech enhancement, audiovisual target speaker extraction, audiovisual speaker separation, CrossNet, AV-CrossNet.
\end{IEEEkeywords}

\section{Introduction}

\IEEEPARstart{I}{n} real-world scenarios, speech signals are often corrupted by acoustic interference, such as background noise, room reverberation, and competing speakers. Such interference significantly degrades speech understanding in both human and computer communications. To mitigate these challenges, speech separation techniques have been developed to separate the target speech signal from interfering sources.
In the past decade, many algorithms have been developed for audio-based speech separation \cite{wang2018supervised}. Traditional approaches 
include speech enhancement that analyzes spectral and statistical properties of speech and noise in order to remove or attenuate background noise \cite{loizou2013speech}, 
and computational auditory scene analysis (CASA) that exploits auditory cues like pitch, onset/offset, and spatial location to segregate sound sources \cite{wang2006computational}.  Since the formulation of speech separation as supervised learning, \glspl{dnn} have become the mainstream approach \cite{wang2018supervised}.

Talker-independent speaker separation faces a so-called permutation ambiguity problem \cite{hershey2016deep}, referring to the challenge of consistently assigning speakers with \gls{dnn} output layers during training. 
This problem can be resolved by deep clustering \cite{kolbaek2017multitalker} or  \glsentrylong{pit} \cite{kolbaek2017multitalker}.
Recent years have seen major progress in speaker separation. Notable methods include Conv-TasNet \cite{luo2019conv}, deep CASA \cite{liu2019divide}, DPRNN \cite{luo2020dprnn}, and TF-GridNet \cite{wang2022tfjournal}.
Despite their impressive results, current speaker separation algorithms have difficulty dealing with audio recordings with partly overlapped speech and same-talker utterances scattered in time \cite{dovrat2021many, von2022segment, taherian2024multi}. 
Target speaker extraction presents an alternative way to address the speaker separation problem. This approach attempts to separate the speech of the target or desired speaker by leveraging discriminative cues that differentiate the target talker from other speakers. Different discriminative cues have been explored such as speaker embedding or identity \cite{wang2018voicefilter}, onset time \cite{pandey2023attentive}, and visual cues \cite{michelsanti2021overview}. 

While existing speaker separation and target speaker extraction methods rely mainly on audio signals, human perception makes use of both auditory and visual cues for speech understanding in multi-talker and adverse acoustic environments \cite{sumby1954visual, crosse2016eye}. Inspired by this multi-modal perception, audiovisual (AV) speech separation aims to incorporate visual information from facial attributes and movements to improve the separation of speech sources \cite{michelsanti2021overview}.

The audio and video modalities exhibit distinct attributes that complement each other for perception. For instance, audio signals are degraded by acoustic interference, which does not impact vision. Visual inputs are affected by poor lighting, limited visual view, and object occlusion, which do not impede audition. By fusing auditory and visual modalities, audiovisual models can achieve more robust separation than audio-only methods, especially in highly noisy environments where auditory cues become unreliable. Additionally, the presence of visual cues may help resolve permutation ambiguity in talker-independent speaker separation.

In recent years, the availability of large-scale audiovisual datasets like LRS2 \cite{afouras2018deep}, VoxCeleb2 \cite{chung2018voxceleb2}, and AVSpeech \cite{ephrat2018looking} has led to the development of promising techniques for 
\gls{avss} \cite{ephrat2018looking, tan2019audio}, 
\gls{avsr} \cite{tamura2015audio, ma2023autoavsr}, 
\gls{avse} \cite{sadeghi2019mixture, li2020visual}, 
and \gls{avtse} \cite{ochiai2019multimodal}. 
Given auditory and visual cues, the task of \gls{avss} is to separate the speech signals of multiple speakers and \gls{avse} aims to separate a single speech signal from non-speech background noise. The goal of \gls{avtse} is to extract the speech signal of a target speaker from a mixture of multiple speakers.

Early studies in audiovisual speaker separation primarily rely on magnitude-domain \gls{tf} masking \cite{afouras2018conversation, afouras2019my, chung2020facefilter, gabbay2018visual, gabbay2018seeing, gao2021visualvoice}. 
%
More recently, time-domain models have been proposed \cite{wu2019time, pan2021muse, ahmadikalkhorani23_interspeech, li2024iianet}, and these 
models employ learnable encoder-decoder modules with demonstrated success in \gls{avss} and \gls{avtse}. Time-domain models usually have more parameters compared to \gls{tf} domain approaches and operate on very short windows of signals for end-to-end speaker separation/extraction.

The introduction of TF-GridNet \cite{wang2022tfjournal}, a complex spectral mapping network for speech separation in single- and multi-channel settings, has renewed interest in \gls{tf} domain models. 
This approach uses larger window lengths and shifts compared to time-domain models, and produces high speech separation performance. 
The audiovisual counterpart of TF-GridNet, AVTF-Gridnet \cite{kalkhorani2024audiovisual}, also shows strong results in \gls{avss} by employing cross-attention audiovisual fusion. 
SAV-GridNet \cite{pan2023scenario}, another audiovisual TF-GridNet model, achieves the best performance in the second COG-MHEAR \gls{avse} challenge in 2023 and substantially outperforms other entries \cite{blanco2023avse}. SAV-GridNet has two scenario-specific models: AV-GridNet${n}$ for \gls{avse} and AV-GridNet${s}$ for \gls{avtse}, and a scenario classifier to guide the system to the model trained for the matched scenario. Despite impressive results, the two expert models trained in their respective scenarios complicate the inference process of AV-GridNet. From the methodological perspective, the use of a scenario classifier commits the system to a particular task early on, and classification errors cannot be corrected by later processing. Moreover, a unified model for \gls{avse} and \gls{avtse} can also enhance learning by leveraging the relationship between the two tasks.

Another recently developed complex spectral mapping model, CrossNet \cite{kalkhorani2024crossnet}, yields comparable performance to TF-GridNet, but with reduced computational expenses. 
CrossNet is an audio-based speech separation model that can be applied to speech enhancement and speaker separation in monaural and multi-channel recordings \cite{kalkhorani2024crossnet}. 
CrossNet has a compact architecture and is easier to train compared to TF-GridNet which utilizes recurrent layers.


In this paper, we propose AV-CrossNet, an audiovisual system based on the CrossNet architecture, for \gls{avse}, \gls{avss}, and \gls{avtse} tasks. We employ a widely adopted visual front-end, DeepAVSR \cite{afouras2018deep}, to extract visual embeddings from video frames. AV-CrossNet includes an early audiovisual fusion layer and positional encoding that preserves the temporal order of audiovisual fused features.  In addition to proposing AV-CrossNet, we demonstrate that it achieves the top performance on multiple trained and untrained \gls{av} datasets. Compared to recently proposed complex spectral mapping models, AV-CrossNet can be trained more efficiently as it has no recurrent layers and employs half-precision training. Another contribution of this paper is that we handle the permutation ambiguity problem in speaker separation by leveraging visual cues.

The rest of this paper is organized as follows. In Section \ref{sec:problem_statement}, we formulate audiovisual speaker separation and target speaker extraction problems. In Section \ref{sec:methodology}, we describe AV-CrossNet in detail. Experimental setup is provided in Section \ref{ssec:experimental_setup}. In Section \ref{sec:results}, we present experimental and comparison results. Section \ref{sec:conclusion} concludes this paper. 

We provide the AV-CrossNet code at \href{https://github.com/ahmadikalkhorani/AVCrossNet}{https://github.com/ahmadikalkhorani/AVCrossNet}.

\section{Problem formulation}
\label{sec:problem_statement}

For a mixture of $C$ speakers in a noisy environment captured by a single microphone, the recorded mixture in the time domain $y(t)$ can be modeled in terms of the clean signals $s_c(t)$ and background noises $n(t)$, as

\nomenclature{$C$}{{number of speakers}\nommodule{Sec \ref{sec:problem_statement}}}
\nomenclature{$y$}{{mixture signal}\nommodule{Sec \ref{sec:problem_statement}}}
\nomenclature{$c$}{{speaker subscript}\nommodule{Sec \ref{sec:problem_statement}}}
\nomenclature{$s_c(t)$}{{clean signal}\nommodule{Sec \ref{sec:problem_statement}}}
\nomenclature{$n(t)$}{{background noises}\nommodule{Sec \ref{sec:problem_statement}}}

\begin{equation}
    y(t) = \sum_{c=1}^C  s_c(t) + n(t),
\end{equation} %
where $t$ denotes discrete time. In the \gls{stft} domain, the model is expressed as:

\begin{equation}
\label{eq:tfdomain}
    Y(m,f)=\sum_{c=1}^C  S_c(m,f) + N(m,f),
\end{equation} where $m$ indexes time frames and $f$ frequency bins. $\mathbf{Y}$, $\mathbf{S}_c$, and $\mathbf{N} \in \mathbb{C}^{M\times F}$ represent the complex spectrograms of the mixture, the clean signal of speaker $c$, and background noise, respectively, and $M$ and $F$ denote the number of time frames and frequency bins, respectively. 
Audiovisual speaker separation based on complex spectral mapping trains a \gls{dnn} to estimate the real and imaginary parts of the signal of each speaker given the mixture $\mathbf{Y}$ and visual stream of each speaker $\mathbf{V}_c \in \mathbb{R}^{M_v \times 3 \times h \times w}$. This is defined as

\nomenclature{$Y$}{{mixture signal in STFT domain}\nommodule{Sec \ref{sec:problem_statement}}}
\nomenclature{$V_c$}{{visual stream of speaker $c$ }\nommodule{Sec \ref{sec:problem_statement}}}
\nomenclature{$\mathbf{S}_c$}{{clean signal in STFT domain}\nommodule{Sec \ref{sec:problem_statement}}}
\nomenclature{$h$}{{face image height }\nommodule{Sec \ref{sec:problem_statement}}}
\nomenclature{$w$}{{face image width}\nommodule{Sec \ref{sec:problem_statement}}}
\nomenclature{$M_v$}{{number of video frames}\nommodule{Sec \ref{sec:problem_statement}}}

\nomenclature{$N$}{{background noises in STFT domain}\nommodule{Sec \ref{sec:problem_statement}}}

\nomenclature{$M$}{{time frames}\nommodule{Sec \ref{sec:problem_statement}}}
\nomenclature{$F$}{{frequency bins}\nommodule{Sec \ref{sec:problem_statement}}}

\nomenclature{$t$}{{time index}\nommodule{Sec \ref{sec:problem_statement}}}
\nomenclature{$f$}{{frequency index}\nommodule{Sec \ref{sec:problem_statement}}}

\nomenclature{$c_T$}{{target speaker subscript}\nommodule{Sec \ref{sec:problem_statement}}}

\begin{equation}
\label{eq:avss}
    \mathbf{\hat{S}}_1, \dots, \mathbf{\hat{S}}_C = \text{DNN}_{\theta}\left(  \mathbf{Y},  \mathbf{V}_1, \dots, \mathbf{V}_C \right), 
\end{equation} %
where $\text{DNN}_{\theta}$ denotes \gls{dnn} parameterized by $\theta$.

In the case of \gls{avtse} and \gls{avse}, $C=1$, and the goal is to estimate the complex spectrogram of the target speaker (or speech signal) $\mathbf{S}_{T}$ from the mixture  $\mathbf{Y}$ and the visual stream of the target speaker $\mathbf{V}_{T}  \in \mathbb{R}^{M_v \times 3 \times h \times w}$.

\section{System description}
\label{sec:methodology}

\begin{figure}[t]
  \centering
  \includegraphics[width=0.999\linewidth]{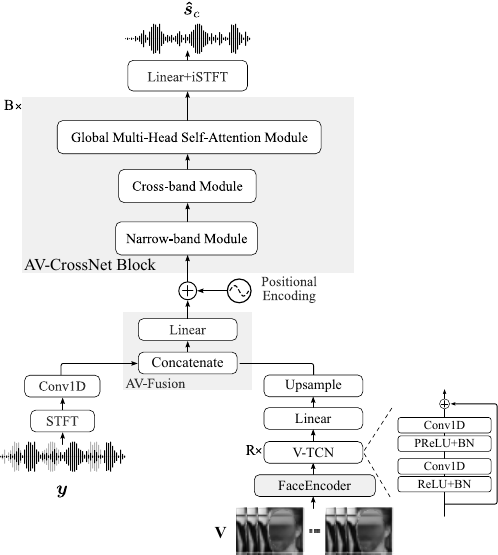}
  \caption{Diagram of the proposed AV-CrossNet system for \gls{avse} and \gls{avtse}. For \gls{avss}, a video component is added for each speaker. Faces in this figure are blurred for privacy reasons.}. 
  \label{fig:AVCrossNet}
\end{figure}

Fig.~\ref{fig:AVCrossNet} shows the diagram of the proposed system. AV-CrossNet comprises an audio encoder layer, a visual encoder, $B$ separator blocks referred to as AV-CrossNet blocks, and a decoder layer. Before processing, we normalize the input signal by its variance to ensure comparable energy levels for all signals processed by AV-CrossNet. We restore the predicted signal's scale using the same variance.
We apply \gls{stft} to the normalized signal, stack the \gls{ri} parts, and send them to the audio encoder layer. This layer learns to extract acoustic features from the input in the \gls{stft} domain. The visual encoder extracts visual features from a whole-face image of a speaker and upsamples the result to match the time resolution of the audio stream.
Audio and visual features are then fused in the AV-fusion block and passed to AV-CrossNet blocks. These blocks are responsible for separating speakers or extracting the target speaker given the fused multi-modal features. Each AV-CrossNet block comprises a narrow-band module, a cross-band module, and a \gls{gmhsa} module. The \gls{gmhsa} module captures global correlations, while the cross-band module captures cross-band correlations. The narrow-band module captures correlations at neighboring frequency bins.
Finally, the decoder layer maps the separated features to a \gls{tf} representation, which is then converted back to the time domain using \gls{istft}.

\subsection{Audio Encoder}
\label{sec:audio_encoder}

\nomenclature{$\mathbf{\hat{Y}}$}{{stacked RI parts}\nommodule{Sec \ref{sec:audio_encoder}}}
\nomenclature{$\Tilde{\mathbf{Y}}$}{{encoded audio}\nommodule{Sec \ref{sec:audio_encoder}}}
\nomenclature{$k_a$}{{audio encoder kernel size}\nommodule{Sec \ref{sec:audio_encoder}}}
\nomenclature{$H$}{{audio encoder output ch.}\nommodule{Sec \ref{sec:audio_encoder}}}

We stack \gls{ri} parts of the input signal in the \gls{stft} domain $\mathbf{\hat{Y}} \in \mathbb{R}^{M \times 2 \times F}$ and feed them to the audio encoder.  
As shown in Fig.~\ref{fig:AVCrossNet}, the audio encoder comprises a convolutional layer with the kernel, stride, and output channel dimension of $k_a$, 1, and $H$, respectively. The audio encoder extends the feature dimension from 2 (\gls{ri}) to $H$. This can be derived as 
\begin{subequations}
\begin{align}
    \hat{\mathbf{Y}} &= [\mathbf{Y}^{(r)}, \mathbf{Y}^{(i)}], &\hat{\mathbf{Y}} \in \mathbb{R}^{M \times 2 \times F}, \\
    \Tilde{\mathbf{Y}} &= \text{Conv1D}(\hat{\mathbf{Y}}),    &\Tilde{\mathbf{Y}} \in \mathbb{R}^{M \times H \times F},
\end{align}
\end{subequations}
where superscripts $(r)$ and $(i)$ indicate real and imaginary parts, respectively, and $[(\cdot), (\cdot)]$ denotes the stacking operation.

\subsection{Visual Encoder}
\label{sec:visual_encoder}

\nomenclature{$R$}{{number of \gls{vtcn} blocks}\nommodule{Sec \ref{sec:visual_encoder}}}

The visual encoder consists of a face encoder, a \gls{vtcn} module, a linear layer, and an up-sampling layer. We assume that all videos are recorded at 25 frames per second and that each talker's face is cropped and scaled to an image with a shape of $1\times112\times112$. The face encoder takes the whole face gray-scaled image $\mathbf{V} \in \mathbb{R}^{M_v \times 1 \times 112 \times 112}$ and converts each frame to an embedding vector yielding $\mathbf{V}' \in \mathbb{R}^{M_v \times F_v}$, where $M_v$ represents the number of video frames and $F_v$ indicates the embedding dimension. \gls{vtcn} \cite{lea2016temporal,luo2019conv} comprises two activation functions and \gls{bn} layers, with each followed by a \gls{conv1d}. The activation functions are \gls{relu} and \gls{prelu}. The input and output channels of both \glspl{conv1d} are equal to the face encoder's output dimension $F_v$. The kernel size of the first and second \gls{conv1d} is 1 and 3, respectively. The \gls{vtcn} module is repeated for $R=5$ times with residual connections within each module.  A linear layer is employed after the \gls{vtcn} blocks, and it converts the feature dimension from $F_v$ to $F$. Finally, the upsampling layer interpolates the feature matrix in the time dimension to match the time resolution $T$ of auditory features. These steps are formulated as follows
\begin{subequations}
\begin{align}
    & \mathbf{V'_c} = \text{FaceEncoder}(V),& \mathbf{V'_c} &\in \mathbb{R}^{M_v \times 1 \times F_v}, \\ 
    & \mathbf{V''_c} = \text{V-TCN}(V'_c),& \mathbf{V''_c} &\in \mathbb{R}^{M_v \times 1 \times F}, \\
    & \mathbf{V'''_c} = \text{Linear}(V''_c),& \mathbf{V'''_c} &\in \mathbb{R}^{M_v \times H \times F}, \\
    & \mathbf{\tilde{V}_c} = \text{Upsample}(V'''_c),& \mathbf{\tilde{V}_c} &\in \mathbb{R}^{M \times H \times F}.
\end{align}
\end{subequations}

\nomenclature{$\tilde{V}$}{{encoded video of single spk}\nommodule{Sec \ref{sec:visual_encoder}}}

In the case of \gls{avss}, $C>1$, and we concatenate the visual features of all speakers along the channel dimension. This is defined as

\begin{align}
    \mathbf{\Tilde{V}} = [\mathbf{\tilde{V}}_1, \mathbf{\tilde{V}}_2, \dots, \mathbf{\tilde{V}}_C],  \qquad \mathbf{\Tilde{V}} \in \mathbb{R}^{M \times (C \times H) \times F}.
\end{align}

\subsection{Audiovisual Fusion}
\label{sec:avfusion}

The audiovisual fusion module, denoted as AV-Fusion in Fig.~\ref{fig:AVCrossNet}, receives auditory $\Tilde{\mathbf{Y}} \in \mathbb{R}^{M \times H \times F}$ and visual features $\mathbf{\Tilde{V}} \in \mathbb{R}^{M \times (C \times H) \times F}$ as input and concatenates them along the second dimension. Then, we use a linear layer to convert the dimension back to $H$. This is defined as

\begin{subequations}
\begin{align}
    & \mathbf{X} = [\Tilde{\mathbf{Y}}, \mathbf{\Tilde{V}}],  & \mathbf{X} &\in \mathbb{R}^{M \times (C\times H+H) \times F}, \\
    & \mathbf{X'} = \text{Linear}(\mathbf{X}), & \mathbf{X'} &\in \mathbb{R}^{M \times H \times F}.
\end{align}
\end{subequations}

\nomenclature{$X$}{{concatenated A and V}\nommodule{Sec \ref{sec:avfusion}}}
\nomenclature{$X'$}{{fused A and V}\nommodule{Sec \ref{sec:avfusion}}}
\nomenclature{$\mathbf{S}_T$}{target speaker's clean signal in the STFT domain\nommodule{Sec \ref{sec:problem_statement}}}
\nomenclature{$\mathbf{V}_T$}{visual stream of the target speaker\nommodule{Sec \ref{sec:problem_statement}}} 

\subsection{Positional Encoding}
\label{sec:rcpe}
To preserve relative positions of the audiovisual features in time, we employ \gls{rcpe} \cite{kalkhorani2024crossnet}, which selects a contiguous chunk of positional embedding vectors from a pre-computed \gls{pe} matrix. For \gls{rcpe}, we start by defining $\text{PE}$ as a combination of sine and cosine functions \cite{vaswani2017attentionisall} as 

\begin{subequations}
    \begin{align}
        \text{{PE}}(m, 2i) =   & \sin\left(\frac{{m}}{{10000^{2i/(F\times H)}}}\right), \\
        \text{{PE}}(m, 2i+1) = & \cos\left(\frac{{m}}{{10000^{2i/(F\times H)}}}\right).
    \end{align}
\end{subequations}
where $i$ indexes the embedding dimension.
Let the fused audiovisual feature matrix have $L$ time frames, where $L=M$ for training and validation, and $L \in [1,L^{\text{max}}]$ for testing,  with $L^{\text{max}} > M$ denoting the maximum desired sequence length in the test set. We define \gls{rcpe} as follows. 
 When the model is in the training mode, we select a random \gls{pe} chunk from frame $\tau$ to $\tau+L-1$, where $\tau$ is drawn randomly from $[1, L^{\text{max}}-L+1]$. When the model is in the test or validation mode, we select the first $L$ embedding vectors, i.e., $\tau = 1$. We obtain positional encoding vectors as shown below
\begin{equation}
    \label{eq:rcpe}
    \text{RCPE}(L) =
    \begin{cases}
        \begin{aligned}
             & \text{PE}[\tau:\tau+L-1, \dots], & \text{if training} \\
             & \text{PE}[1:L, \dots],   & \text{else}
        \end{aligned}
    \end{cases}
\end{equation}

It has been shown that \gls{rcpe} generalizes better for out-of-distribution sequence lengths \cite{kalkhorani2024crossnet}.
Additionally, \gls{rcpe} has no learnable parameter and has a negligible computational cost.  Finally, we reshape and add the selected \glspl{pe} to input features:
\begin{align}
    & \mathbf{X''} = \mathbf{X'} + \text{Reshape}(\text{RCPE}(L)).
\end{align}
\nomenclature{$RCPE$}{{random chunk positional encoding}\nommodule{Sec \ref{sec:rcpe}}}
\nomenclature{$X''$}{{fused feature + RCPE}\nommodule{Sec \ref{sec:rcpe}}}
\nomenclature{$L^{\text{max}}$}{{max test frame length }\nommodule{\gls{rcpe}}}
\nomenclature{$\tau$}{{random time index}\nommodule{\gls{rcpe}}}
\nomenclature{$PE$}{{positional embedding matrix}\nommodule{\gls{rcpe}}}

\subsection{AV-CrossNet Block}

\begin{figure*}
    \centering
    \includegraphics[width=0.98\linewidth]{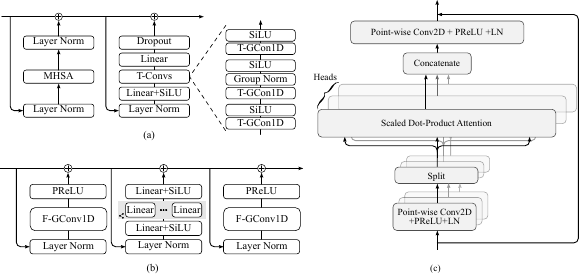}
    \caption{AV-CrossNet building blocks. (a) Narrow-band module.  (b) Cross-band module. (c) Global multi-head self-attention module.}
    \label{fig:tf-crossnet-modules}
\end{figure*}

\subsubsection{Narrow-band Module}

The narrow-band module, as illustrated in Fig.~\ref{fig:tf-crossnet-modules}(a), comprises two components: \gls{mhsa} and \gls{tconv}. The \gls{mhsa} component is encompassed by two \gls{ln} layers and operates on the input feature matrix $\mathbf{X''}$. The output of the \gls{mhsa} component is subsequently added to the input of the narrow-band module. Notably, this module can function as an additional audiovisual fusion layer, similar to an attentive audiovisual fusion mechanism \cite{kalkhorani2024audiovisual, ahmadikalkhorani23_interspeech}.
The \gls{tconv} component consists of the following layers: an \gls{ln} layer, a linear layer with a subsequent \gls{silu} activation, a \gls{tconv} layer, and a final linear layer. The initial linear layer within this component increases the feature dimension in the input from $H$ to $H''$, while the last linear layer reverts the feature dimension back to $H$. Finally, we obtain the output of the  narrow-band module by passing the result to a Dropout layer and adding the module's input.

\subsubsection{Cross-band Module}

To model cross-band correlations in the input signal, we utilize the cross-band module introduced in \cite{quan2023spatialnet, kalkhorani2024crossnet}. This module, depicted in Fig.~\ref{fig:tf-crossnet-modules}(b), incorporates two frequency-convolutional components and a full-band linear component. The frequency-convolutional components are designed to capture correlations among neighboring frequencies. Each comprises an \gls{ln} layer, a grouped convolution layer along the frequency axis (F-GConv1d), and a \gls{prelu} activation function. In the full-band linear component, we initially employ a linear layer followed by a \gls{silu} activation function to reduce the number of hidden channels from $H$ to $H'$. Subsequently, a series of linear layers along the frequency axis is applied to capture full-band features. Each feature channel is associated with a dedicated linear layer denoted as $\text{Linear}_i$ for $i=1, \ldots, H'$, as illustrated in Fig.~\ref{fig:tf-crossnet-modules}(b). Notably, the parameters of these linear layers are shared among all AV-CrossNet blocks, similar to \cite{kalkhorani2024crossnet}. Finally, the output of the component is obtained by increasing the number of channels back to $H$ using a linear layer with \gls{silu} activation and adding it to the original input of this module.

\subsubsection{Global Multi-Head Self-Attention Module}

Figure \ref{fig:tf-crossnet-modules}(c) illustrates the \gls{gmhsa} module. A convolutional layer with $L(2E+H/L)$ output channels extracts frame-level features from each \gls{tf} unit. The output is then divided into $L$ queries $Q^l\in\mathbb{R}^{E \times F \times T}$, keys $K^l\in\mathbb{R}^{E \times F \times T}$, and values $V^l \in\mathbb{R}^{H/L \times F \times T}$, where $E$ represents the output channel dimension of the point-wise convolution, and $l$ indexes the head number. We apply \gls{mhsa} to these embeddings to capture global dependencies. The outputs from all heads are concatenated and processed by another point-wise convolution with an output dimension of $H$, followed by a \gls{prelu} activation and an \gls{ln} layer. The resulting values are added to the input of the \gls{gmhsa} to produce the final output.

\subsection{Decoder} 

As shown in Fig. \ref{fig:AVCrossNet}, the decoder uses a linear layer to convert the feature dimension from $H$ to $2\times C$ resulting in $Z \in \mathbb{R}^{C\times 2 \times M \times F}$, which is then converted to $C$ waveform signals using \gls{istft}.

\subsection{Loss Function}
We employ a combination of magnitude loss, $\mathcal{L}_{\text {Mag}}$, and \gls{si-sdr} loss, $\mathcal{L}_{\text {SI-SDR}}$, similar to \cite{wang2022tfjournal,pan2022hybrid}. We use the standard form of \gls{si-sdr} where the target signal is scaled to match the scale of the estimated signal. Also, we scale the magnitude loss by the  $L_1$ norm of the magnitude of the target signal in the \gls{stft} domain similar to \cite{pan2022hybrid}. These loss functions are defined as

\begin{subequations}
    \label{eq:loss}
    \begin{align}
        & \mathcal{L}= \mathcal{L}_{\text{Mag}} + \mathcal{L}_{\text{SI-SDR}}, \\
        & \mathcal{L}_{\text{Mag}} = \frac{\left\| | \operatorname{STFT}(\hat{\boldsymbol{s}}_c) | - | \operatorname{STFT}(\boldsymbol{s}_c) | \right\|_1}{\left\| | \operatorname{STFT}(\boldsymbol{s}_c) | \right\|_1}, \\
        & \mathcal{L}_{\text{SI-SDR}} = -\sum_{c=1}^C 10 \log_{10} \frac{\|\boldsymbol{s}_c\|_2^2}{\|\hat{\boldsymbol{s}}_c - \alpha_c \boldsymbol{s}_c\|_2^2}, \\
        & \alpha_c = \frac{\boldsymbol{s}_c^T \hat{\boldsymbol{s}}_c}{\boldsymbol{s}_c^T \boldsymbol{s}_c}.
    \end{align}
\end{subequations}

In the above equations, $L_1$ norm is denoted by $\left\|\cdot\right\|_1$, $|\cdot|$ represents the magnitude operator, $\alpha_c$ is the scaling factor, and $(\cdot)^T$ denotes the transpose operation. 
To resolve permutation ambiguity during the \gls{avss} training, we order the target signals in the same order as the visual streams. Thus the model learns to map each talker's signal to the corresponding output layer consistently. 
\section{Experimental Setup}
\label{ssec:experimental_setup}

\subsection{Data Preparation}

We evaluate AV-CrossNet on four tasks, i.e. \glsentrylong{avtse}, \glsentrylong{avss}, \glsentrylong{avss} in noisy environments, and \glsentrylong{avse}.

\subsubsection{Audiovisual Speaker Separation in Clean Environments} 

We use LRS2 \cite{afouras2018deep}, LRS3 \cite{afouras2018lrs3}, VoxCeleb2 \cite{chung2018voxceleb2} datasets for evaluating \gls{avss}. LRS2 contains thousands of single-talker BBC video clips. We derive the LRS2-2Mix dataset from LRS2 by mixing different speakers with \gls{tir} ranging from -5 to +5~dB. The training set comprises 11 hours of mixed audio, while the validation set has 3 hours. The test set is the same one used in prior studies \cite{li2022audio}. LRS3 consists of thousands of spoken sentences from TED and TEDx videos. To create the LRS3-2Mix dataset, we mix various speakers in the \texttt{Trainval} subset, ensuring that there is no overlap between the training and validation sets. We follow the same \gls{tir} range as in LRS2-2Mix.

\subsubsection{Audiovisual Speaker Separation in Noisy Environments}

We use the NTCD-TIMIT dataset \cite{abdelaziz2017ntcd}, recorded in controlled environments, and the LRS3+WHAM! dataset. Both are pre-processed to extract the lip region from the video recordings, following the approach suggested in \cite{afouras2018deep}. The NTCD-TIMIT dataset is designed for noise-robust audiovisual speech recognition for single speakers. To create mixtures of two speakers with noise, we randomly select a second speaker from the same dataset, ensuring that no overlap exists in speakers or utterance contents. \gls{tir} is set to 0~dB for speech signals, while the noise signal is uniformly sampled in the \gls{snr} range $[-5,~20]$~dB with mixed signal considered as signal. The dataset has 39 training, 8 validation, and 9 test speakers, with 5 hours of training data, 1 hour of validation data, and 1 hour of test data.

The LRS3+WHAM! dataset combines audiovisual data from LRS3 \cite{afouras2018lrs3} with noise data from WHAM! \cite{wichern2019wham}. We generate synthetic mixtures of two speakers plus noise. The \gls{tir} of clean speech mixtures is uniformly sampled in the range of $[-5,~5]$~dB, and \gls{snr} is in the range of $[-6,~3]$~dB, following the original mixing process from WHAM!. The LRS3+ WHAM! dataset provides 28 hours of training data, 3 hours of validation data, and 2 hours of test data.

\subsubsection{Audiovisual Target Speaker Extraction} 

To assess AV-CrossNet's performance on the \gls{avtse} task, we utilize the VoxCeleb2 \cite{chung2018voxceleb2} dataset for training. This dataset contains over one million utterances from YouTube videos, in a variety of video resolutions and acoustic environments. In order to build the training, validation, and test datasets, we ensure that speakers in the training dataset are not included in the test dataset and all utterances are longer than four seconds. For the training, validation, and test datasets, we generate 20,000, 5,000, and 3,000 mixtures of two speakers, respectively. We introduce a competing speaker to the target speech by randomly sampling a \gls{tir} in the range of -10 to 10~dB. In addition to the VoxCeleb2 dataset, we examine AV-CrossNet's performance on two other untrained datasets, LRS3 \cite{afouras2018lrs3} and TCD-TIMIT \cite{harte2015tcd}. In each case, we generate 3,000 mixtures similar to \cite{pan2021muse, wu2019time, lin2023avsepformer}.

\subsubsection{Audiovisual Speech Enhancement}

We use the second COG-MHEAR challenge dataset \cite{blanco2023avse} for \gls{avse} and further \gls{avtse} evaluations. The speech signals in this dataset are collected from the LRS3 dataset \cite{afouras2018lrs3}. The noise signals are collected from three sources, the domestic noises from the Clarity challenge \cite{graetzer2021clarity}, the Demand noise set \cite{thiemann2013demand}, and the Freesound noise set from the second DNS challenge \cite{dubey2023deep}. From the COG-MHEAR training set, we select the utterances of 24 speakers with the largest number of utterances to create the validation dataset, resulting in 1339 utterances for \gls{avse} and 1358 utterances for \gls{avtse}. The remaining ones form the training dataset. We treat the development set as the test dataset. The number of training, validation, and test utterances are 30,297, 3,017, and 3,306. The lengths of both the training and validation utterances are 3.0 seconds. Test utterance lengths range from around two seconds to over thirty seconds. This dataset adopts speech-weighted \glspl{snr} which are randomly drawn from [-10,~10]~dB and [-15,~5]~dB, for \gls{avse} and \gls{avtse} respectively.

\subsection{DNN Configuration}
\label{sec:network_config}

\nomenclature{$B$}{{number of CrossNet blocks}\nommodule{Fig. 1 and Sec \ref{sec:network_config}}}

For the proposed AV-CrossNet architecture, we make use of the hyperparameters in \cite{kalkhorani2024crossnet}. We set the kernel size of the audio encoder layer $k_a$, time-dimension group convolution (T-GConv1d), and frequency-dimension group convolution (F-GConv1d) to 5, 5, and 3, respectively. The number of groups for T-GConv1d, F-GConv1d, and group normalization is all set to 8. The proposed model comprises $B = 12$ blocks, with hidden channel sizes set to $H = 192$, $H' = 16$, and $H'' = 384$. We employ $N=4$ self-attention heads in the \gls{gmhsa} module with an embedding dimension $E=\left\lceil 512/F \right\rceil$, where $\left\lceil \cdot \right\rceil$ denotes ceiling operation.

To process the input data, we apply \gls{stft} using a Hanning window with frame length of 512 samples ({32~ms}) and frame shift of 256 samples ({16~ms}).
We employ $R=5$ \gls{vtcn} blocks, where the convolutional layers utilize a stride of 1. The output from the \gls{vtcn} blocks is then passed into a \gls{conv1d} layer with a kernel size of 3 and 512 output channels.

We utilize the Adam optimization algorithm with a maximum learning rate of $10^{-3}$. Initially, we utilize a cosine warm-up scheduler that gradually increases the learning rate from $10^{-6}$ to $10^{-3}$ over the first 10 epochs, similar to the approach described in \cite{kalkhorani2024crossnet}. Subsequently, we transition to the PyTorch ReduceLROnPlateau scheduler, setting the patience to 3 epochs and the reduction factor to 0.9. To reduce memory requirements and accelerate training, we adopt the half-precision (mixed-16) training strategy in our experiments. The model training continues until the validation loss fails to improve for 10 consecutive epochs. For each experiment, we utilize the maximum number of batches that can fit into the GPU memory (NVIDIA A100 GPU with 40 GB of memory).

\subsection{Evaluation Metrics}

To evaluate AV-CrossNet's performance, we use several common objective metrics. These include: \gls{si-sdr} and its improvement, SI-SDRi \cite{le2019sdr}, SDR and its improvement \cite{vincent2006performance}, scale-invariant SNR, narrow- and wide-band \gls{pesq} \cite{rix2001pesq}, and \gls{estoi} \cite{jensen2016algorithmESTOI}.

To compute these metrics, we employ the TorchMetrics[audio] package \cite{detlefsen2022torchmetrics}, which provides a set of evaluation tools designed specifically for audio-related tasks. For calculating narrow-band \gls{pesq}, we use the \textit{pypesq} package\footnote{https://github.com/vBaiCai/python-pesq} as done in \cite{li2022audio,li2024iianet}.

\section{Experimental Results and Comparisons} 
\label{sec:results} 

\subsection{Audiovisual Speaker Separation in Clean Environments}

\begin{table*}[ht]
  \renewcommand\arraystretch{1.2}

  \caption{Speaker separation results of different AVSS methods on LRS2, LRS3, and VoxCeleb2 datasets. "Mod." stands for modality, which is either audio-only (A) or audiovisual (AV). The results of comparison methods are adopted from \cite{li2022audio} and \cite{li2024iianet}.}
  \centering

  \begin{tabularx}{0.99\linewidth}{Xccccccccccccc}
    \toprule
    \multirow{2}{*}{Method}                         &            &        &      & \multicolumn{3}{c}{LRS2} & \multicolumn{3}{c}{LRS3} & \multicolumn{3}{c}{VoxCeleb2 }                                                                   \\
    \cmidrule(r){5-7} \cmidrule(r){8-10} \cmidrule(r){11-13}
                                                    & Params (M) & Domain & Mod. & SI-SNRi                  & SDRi                     & PESQ                           & SI-SNRi  & SDRi     & PESQ     & SI-SNRi  & SDRi     & PESQ     \\
    \midrule
    Unprocessed                                     &            &        &      & 0.0                      & 0.0                      & 1.71                           & 0.0      & 0.0      & 1.73     & 0.0      & 0.0      & 1.89     \\
    \midrule
    DPCL++ \cite{hershey2016deep}                   & 13.6       & T-F    & A    & 3.3                      & 4.3                      & --                             & 5.8      & 6.2      & --       & 2.1      & 2.5      & --       \\
    Conv-TasNet \cite{luo2019conv}                  & 5.6        & Time   & A    & 10.3                     & 10.7                     & --                             & 11.1     & 11.4     & --       & 6.9      & 7.5      & --       \\
    SuDoRM-RF  \cite{tzinis2020sudo}                & 2.7        & Time   & A    & 9.1                      & 9.5                      & --                             & 12.1     & 12.3     & --       & 6.5      & 6.9      & --       \\
    A-FRCNN \cite{hu2021speech}                     & 6.3        & Time   & A    & 9.4                      & 10.1                     & --                             & 12.5     & 12.8     & --       & 7.8      & 8.2      & --       \\
    \midrule
    The Conversation \cite{afouras2018conversation} & 62.7       & T-F    & AV   & --                       & --                       & --                             & --       & --       & --       & --       & 8.9      & --       \\
    AVConvTasNet  \cite{wu2019time}                 & 16.45      & Time   & AV   & 12.5                     & 12.8                     & 2.69                           & 11.2     & 11.7     & 2.58     & 9.2      & 9.8      & 2.17     \\
    LWTNet  \cite{afouras2020self}                  & 69.4       & T-F    & AV   & --                       & 10.8                     & --                             & --       & 4.8      & --       & --       & --       & --       \\
    Visualvoice  \cite{gao2021visualvoice}          & 77.8       & T-F    & AV   & 11.5                     & 11.8                     & 2.78                           & 9.9      & 10.3     & 2.13     & 9.3      & 10.2     & 2.45     \\
    CaffNet-C \cite{lee2021looking}                 &            & T-F    & AV   & --                       & 10.0                     & --                             & --       & 9.8      & --       & --       & 7.6      & --       \\
    CTCNet  \cite{li2022audio}                      & 7.0        & Time   & AV   & 14.3                     & 14.6                     & 3.08                           & 17.4     & 17.5     & 3.24     & 11.9     & 13.1     & 3.00     \\
    AVLiT   \cite{martel2023audio}                  & 5.75       & Time   & AV   & 12.8                     & 13.1                     & 2.56                           & 13.5     & 13.6     & 2.78     & 9.4      & 9.9      & 2.23     \\
    IIANet \cite{li2024iianet}                      & 3.1        & Time   & AV   & 16.0                     & 16.2                     & 3.23                           & \bf 18.3 & \bf 18.5 & 3.28     & 13.6     & 14.3     & 3.12     \\
    AVSepChain \cite{mu2024separate}                & 33.1       & Time   & AV   & 15.3                     & 15.7                     & 3.26                           & --       & --       & --       & 13.6     & 14.2     & 2.72     \\
    \midrule
    \bf AV-CrossNet                                 & 11.1       & T-F    & AV   & \bf 16.8                 & \bf 17.1                 & \bf 3.56                       & \bf 18.3 & \bf 18.5 & \bf 3.67 & \bf 14.6 & \bf 14.9 & \bf 3.41 \\

    \bottomrule
  \end{tabularx}

  \label{tab:avss}
\end{table*}

Table~\ref{tab:avss} presents AV-CrossNet's \gls{avss} results in clean environments and compares to those of strong comparison approaches on three LRS2, LRS3, and VoxCeleb2 datasets. Results are reported in terms of SI-SNRi, SDRi, and \gls{pesq}. The table also provides the number of trainable parameters and domain of operation for each model. 

On LRS2, AV-CrossNet obtains an SI-SNRi of 16.8~dB, outperforming the recently-posted, previous best IIANet by 0.8~dB. Its SDRi of 17.1~dB improves upon IIANet's 16.2~dB by 0.9~dB. AV-CrossNet's \gls{pesq} of 3.56 represents a large improvement over IIANet's 3.23. Clear from the table, there is a considerable performance gap between the best AV and audio-only models on all datasets, demonstrating the contribution of the visual stream. On LRS3, AV-CrossNet matches the 18.3~dB SI-SNRi and 18.5~dB SDRi achieved by IIANet, but exceeds IIANet's \gls{pesq} score by a large margin of 0.39, reaching 3.67 \gls{pesq}. 

VoxCeleb2 is a more challenging dataset, and AV-CrossNet achieves the state-of-the-art scores of 14.6~dB SI-SNRi, 14.9~dB SDRi, and 3.41 \gls{pesq}. These results outperform the previous best-performing model of IIANet by 1.0~dB SI-SNRi, 0.6~dB SDRi, and 0.29 \gls{pesq}. 

As described earlier, AV-CrossNet does not explicitly address the permutation ambiguity issue. Instead we exploit the fact that the visual stream only contains one speaker, and hence it does not pose a permutation issue. In other words, we leverage audiovisual integration to resolve the permutation ambiguity problem. The above compelling separation results demonstrate that permutation ambiguity can be entirely avoided in audiovisual speaker separation.

\subsection{Audiovisual Speaker Separation in Noisy Environments} 

Table~\ref{tab:avss-noisy} reports AV-CrossNet's \gls{avss} performance in noisy environments and compares to several audio-only and audiovisual methods on two benchmark datasets: NTCD-TIMIT and LRS3+WHAM!. Results are reported in terms of \gls{pesq}, \gls{estoi}, and \gls{si-sdr}i.
On the NTCD-TIMIT dataset, AV-CrossNet achieves the best scores across all three metrics, substantially outperforming current state-of-the-art models. For example, our model obtains the \gls{pesq} value of 2.16, a large improvement over the previous best VisualVoice with 1.45 \gls{pesq}. For \gls{estoi}, AV-CrossNet obtains 0.62, exceeding AVLiT's 0.45 by 0.17. AV-CrossNet's \gls{si-sdr}i of 15.6~dB represents a 4.6~dB gain over AVLiT. Compared to the top audio-only models of DPRNN the improvements are even more pronounced.

On the more challenging LRS3+WHAM! dataset, AV-CrossNet maintains the best performance with a \gls{pesq} of 2.03, surpassing AVLiT by 0.51. Its \gls{estoi} of 0.77 improves upon AVLiT's 0.68 by 0.09. For \gls{si-sdr}i, AV-CrossNet achieves 13.84~dB, a 1.42~dB increase over AVLiT's previous best of 12.42~dB. The best audio-only model of DPRNN scores 10.93~dB \gls{si-sdr}i.
These consistent and often large performance improvements demonstrate AV-CrossNet's superior ability for speech separation and enhancement across different datasets and metrics.

\begin{table}[t]
\renewcommand\arraystretch{1.2}
  \setlength{\tabcolsep}{2.0pt}
  \caption{Speaker separation results of AV-CrossNet and baseline models on NTCD-TIMIT and LRS3+WHAM! datasets. Comparison results are from \cite{martel2023audio}. }
  \label{tab:baseline_quality_comparison}
  \centering

  \begin{tabularx}{0.999\linewidth}{X   c   ccc  ccc }
    \toprule
    \multicolumn{1}{c}{} &  & \multicolumn{3}{c}{NTCD-TIMIT} & \multicolumn{3}{c}{LRS3+WHAM!}
    \\
    \cmidrule(r){3-5}\cmidrule(r){6-8}
    Method                                & Mod.       & PESQ & eSTOI & SI-SDRi & PESQ & eSTOI & SI-SDRi \\
    \hline
    Unprocessed                           &            & 1.19 & 0.33  & --      & 1.08 & 0.37  & --      \\
    \hline
    ConvTasNet \cite{luo2019conv}         & A          & 1.35 & 0.38  & 8.76    & 1.24 & 0.51  & 9.66    \\
    DPRNN \cite{luo2020dual}              & A          & 1.32 & 0.39  & 9.31    & 1.40 & 0.45  & 10.93   \\
    A-FRCNN \cite{hu2021speech}           & A          & 1.30 & 0.31  & 6.92    & 1.25 & 0.49  & 9.21    \\
    \hline
    AVConvTasNet \cite{wu2019time}        & AV         & 1.33 & 0.40  & 9.02    & 1.29 & 0.60  & 6.21    \\
    LAVSE \cite{chuang2020lite}           & AV         & 1.31 & 0.37  & 6.22    & 1.24 & 0.50  & 5.59    \\
    L2L \cite{ephrat2018looking}          & AV         & 1.23 & 0.26  & 3.36    & 1.16 & 0.51  & 7.60    \\
    VisualVoice \cite{gao2021visualvoice} & AV         & 1.45 & 0.43  & 10.04   & 1.48 & 0.63  & 11.87   \\
    AVLiT \cite{martel2023audio}          & AV         & 1.43 & 0.45  & 11.00   & 1.52 & 0.68  & 12.42   \\
    \hline 
    
    \bf AV-CrossNet                   & AV         & \bf 2.16 & \bf 0.62  & \bf 15.6    &  \bf 2.03    & \bf 0.77   & \bf 13.84       \\
    
    \bottomrule
  \end{tabularx}

  \label{tab:avss-noisy}

\end{table}

\subsection{Audiovisual Target Speaker Extraction} 

Table~\ref{tab:avtse} reports AV-CrossNet's \gls{avtse} performance and those of several recent audiovisual methods on three benchmark datasets: LRS3, TCD-TIMIT, and VoxCeleb2. 
Results are given in terms of \gls{si-sdr} and \gls{pesq}.
All models in Table~\ref{tab:avtse} utilize the same visual embedding extractor and are trained only on VoxCeleb2, and tested on LRS3, TCD-TIMIT, and VoxCeleb2 datasets to assess their generalization capabilities.

On the LRS3 dataset, AV-CrossNet achieves the state-of-the-art SI-SDR of 17.42~dB, outperforming the recently posted previous best AV-SepChain by 2.22~dB. Its \gls{pesq} of 3.14 represents a marginal improvement of 0.02 over AV-SepChain.
Similarly, on TCD-TIMIT, AV-CrossNet achieves the best results with 18.15~dB SI-SDR and 3.25 \gls{pesq}, showing an improvement of 3.45~dB SI-SDR and 0.37 in \gls{pesq} over AV-SepChain. 
On the VoxCeleb2 dataset used for training, AV-CrossNet again outperforms all previous methods, achieving 14.71~dB SI-SDR and 2.93 \gls{pesq}. The improvements over AV-SepChain are 0.51~dB in SI-SDR and 0.21 in \gls{pesq}, highlighting the consistently high performance of AV-CrossNet in leveraging audiovisual cues for improved speech separation. Overall, these results establish AV-CrossNet as the new leading method for audiovisual target speaker extraction across diverse audiovisual datasets.

\begin{figure}[ht]
  \centering
  \includegraphics[width=0.99\linewidth]{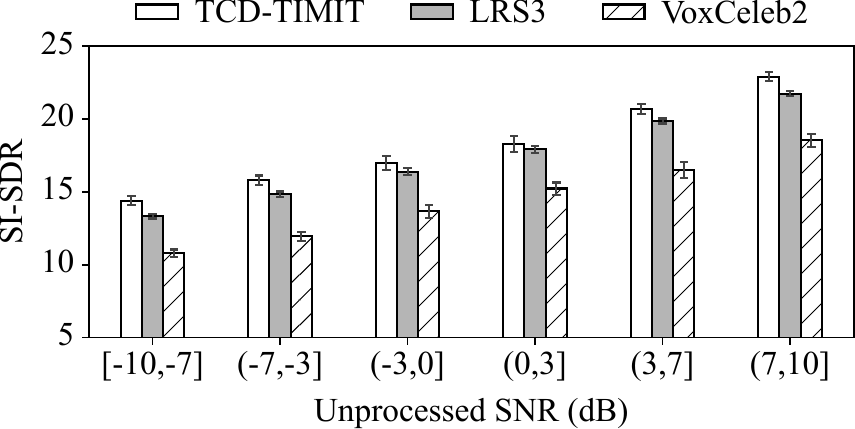}
  \caption{SI-SDR results of AV-CrossNet across different SNRs for unprocessed mixtures on the test datasets TCD-TIMIT, LRS3, and VoxCeleb2. \textcolor{black}{Error bars represent 95\% confidence intervals.}}
  \label{fig:si-sdr-performance}
\end{figure}

In Fig.~\ref{fig:si-sdr-performance}, we break down our \gls{si-sdr} results on three datasets in different ranges of unprocessed SNR (dB). As illustrated in this figure, AV-CrossNet produces excellent results over all input SNR range.  It is worth noting that, as the input SNR range varies over 17~dB -- from [-10,~-7] to (7,~10] -- the processed SI-SDR results vary within a considerably smaller range, about half of 17~dB. This shows that AV-CrossNet yields relatively better performance at lower SNRs, likely due to the more prominent role of the visual modality with decreasing SNR.

\begin{table}
  \renewcommand\arraystretch{1.2}
  \setlength{\tabcolsep}{3.0pt}
  \centering
  \caption{ Target speaker extraction results of AV-CrossNet and baseline models on LRS3, TCD-TIMIT, and VoxCeleb2. All models use the same visual embedding extractor and are trained on the VoxCeleb2 dataset.}
  \begin{tabularx}{0.99\linewidth}{Xcccc|cc}
    \toprule
    & \multicolumn{2}{c}{ LRS3 } &  \multicolumn{2}{c}{ TCD-TIMIT }& \multicolumn{2}{c}{ VoxCeleb2 } \\
    \cmidrule(r){2-3}\cmidrule(r){4-5}\cmidrule(r){6-7}
    Method                                 & SI-SDR    & PESQ     & SI-SDR    & PESQ     & SI-SDR    & PESQ     \\
    \hline
    Unprocessed                            & 0.13      & 1.21     & -0.15     & 1.47     & -0.08     & 1.24     \\
    \hline
    VisualVoice \cite{gao2021visualvoice}  & 11.60     & 2.27     & 10.88     & 2.25     & 9.73      & 1.97     \\
    AV-ConvTasNet \cite{wu2019time}        & 12.13     & 2.33     & 11.53     & 2.21     & 10.38     & 1.97     \\
    MuSE \cite{pan2021muse}                & 12.97     & 2.56     & 12.50     & 2.45     & 11.24     & 2.20     \\
    AV-SepFormer \cite{lin2023avsepformer} & 13.81     & 2.67     & 13.44     & 2.57     & 12.13     & 2.31     \\
    AVSepChain \cite{mu2024separate}       & 15.2      & 3.12     & 14.7      & 2.88     & 14.2      & 2.72     \\
    \hline
    \bf AV-CrossNet                        & \bf 17.42 & \bf 3.14 & \bf 18.15 & \bf 3.25 & \bf 14.71  & \bf 2.93 \\
    \bottomrule
  \end{tabularx}
  \label{tab:avtse}
\end{table}

\begin{figure}[h]
  \centering
  \includegraphics[width=\linewidth]{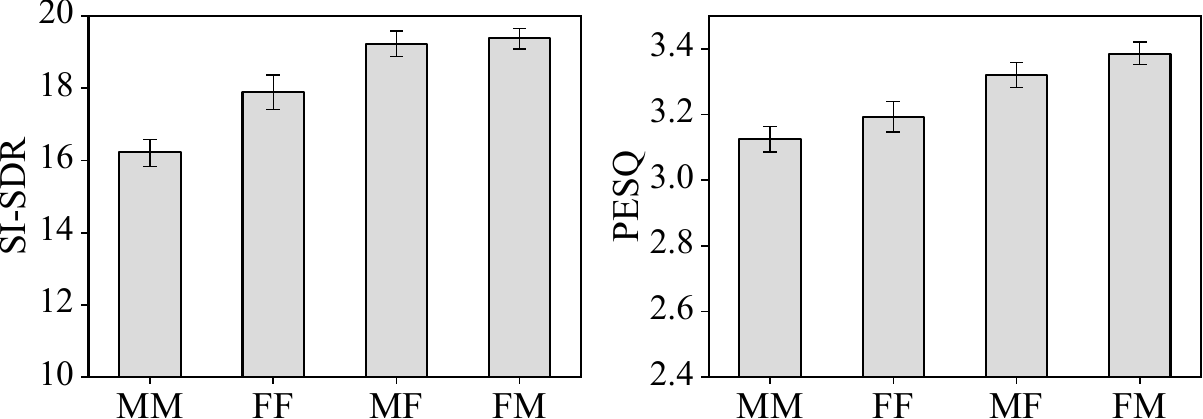}
  \caption{Target speaker extraction results of different gender pairs. The first and second letters denote the genders of the target and interfering speakers, respectively.}
  \label{fig:genderattribute}
\end{figure}

To gain further insight into AV-CrossNet's target speaker extraction performance, we analyze test results on the untrained TCD-TIMIT dataset \cite{harte2015tcd}. Fig.~\ref{fig:genderattribute} breaks down results in various gender pairs. A pair is denoted in the order of target and interferer; for example, MF denotes a male target speaker mixed with a female interfering speaker. We select \gls{si-sdr} and \gls{pesq} metrics to present our analysis. We observe that AV-CrossNet performs better in mixed-gender pairs, not surprisingly. FM pairs are easier to extract than MF pairs. This could be attributed to the higher fundamental frequencies of female speakers, which lead to neighboring harmonics more separated in frequency, facilitating their extraction. In same-gender scenarios, FF pairs yield better results than MM pairs. These findings are generally consistent with the previous findings such as those in \cite{ephrat2018looking}.

\subsection{COG-MHEAR AVSE Challenge}

Table~\ref{tab:avsec2} presents AV-CrossNet's result and compares to the previous state-of-the-art methods on the COG-MHEAR challenge. AV-DPRNN \cite{pan2023scenario} is a time-domain audiovisual model that performs end-to-end audiovisual processing, and is jointly trained on both \gls{avse} and \gls{avtse} for this challenge. AV-GridNet also performs complex spectral mapping, and its results are reported in several training setups. AV-GridNet$_{all}$ is trained in both \gls{avse} and \gls{avtse} scenarios. AV-GridNet$_{s}$ and AV-GridNet$_{n}$ are expert models trained separately for \gls{avtse} and \gls{avse}, respectively. SAV-GridNet denotes the best-performing system on COG-MHEAR challenge that includes two expert models and a scenario classifier. This model has around 22M parameters, aiming to process noisy speech in matched scenarios. 
As shown in Table~\ref{tab:avsec2}, AV-CrossNet with around 11M parameters outperforms AV-GridNet$_{all}$ by 0.13 \gls{pesq} and at least 0.4~dB \gls{si-sdr} on both \gls{avse} and \gls{avtse} tasks.
Notably, AV-CrossNet also outperforms the more sophisticated two-stage system of SAV-GridNet by a substantial margin on \gls{avse}; AV-CrossNet and SAV-GridNet achieve comparable results on the \gls{avtse} task. \textcolor{black}{We note that, although the scenario classifier achieves the classification accuracy of 99.4\% on this particular task}, distinguishing speech enhancement and target speaker extraction is likely error-prone in realistic environments that contain both overlapped speech and background noise. The performance gap between AV-GridNet$_{all}$ and AV-CrossNet shows the latter's versatility in both scenarios.

\begin{table}[h]
  \caption{Evaluation results of AV-CrossNet and baseline models on the development set for the second COG-MHEAR Audiovisual Speech Enhancement Challenge. }

  \renewcommand\arraystretch{1.2}
  \setlength{\tabcolsep}{1.0pt}

  \begin{tabularx}{\linewidth}{Xcccccccc}
    \toprule
                                              & \multicolumn{3}{c}{Speech+Noise} &          & \multicolumn{3}{c}{Speech+Speech}                                     \\
    \cline{2-4} \cline{6-8}
    Method                                    & PESQ                             & STOI     & SI-SDR                            &  & PESQ     & STOI     & SI-SDR   \\
    \midrule
    Unprocessed                               & 1.15                             & 0.68     & -4.4                              &  & 1.17     & 0.60     & -5.0     \\
    \midrule
    AV-DPRNN \cite{pan2023scenario}           & 2.02                             & 0.86     & 11.4                              &  & 2.23     & 0.90     & 12.6     \\
    AV-GridNet$_{all}$ \cite{pan2023scenario} & 2.62                             & 0.91     & 13.9                              &  & 3.10     & \bf 0.95 & 16.7     \\
    AV-GridNet$_s$ \cite{pan2023scenario}     & 2.56                             & 0.90     & 13.4                              &  & \bf 3.23 & \bf 0.95 & \bf 17.5 \\
    AV-GridNet$_n$ \cite{pan2023scenario}     & 2.68                             & 0.91     & 14.2                              &  & 1.27     & 0.61     & -4.7     \\
    SAV-GridNet \cite{pan2023scenario}        & 2.68                             & 0.91     & 14.2                              &  & \bf 3.23 & \bf 0.95 & \bf 17.5 \\
    \midrule
    \textbf{AV-CrossNet}                      & \bf 2.75                         & \bf 0.92 & \bf 14.3                          &  & \bf 3.23 & \bf 0.95 & 17.3     \\
    \bottomrule
  \end{tabularx}

  \label{tab:avsec2}
\end{table}

\section{Conclusion}
\label{sec:conclusion}

We have presented AV-CrossNet for single-channel audiovisual speech enhancement, audiovisual speaker separation, and audiovisual target speaker extraction tasks. This system is based on the CrossNet architecture and integrates a visual front-end using the DeepAVSR model to extract embeddings from video frames. 
AV-CrossNet comprises an audio encoder, a visual encoder, separator blocks, and a decoder. The proposed model leverages narrow- and cross-band modeling as well as global attention to estimate the real and imaginary spectrogram of an individual speaker.
By virtue of audiovisual processing, AV-CrossNet is not susceptible to the permutation ambiguity problem in talker-independent speaker separation.
The proposed model achieves superior performance across a number of audiovisual datasets, even on untrained challenging datasets, and establishes new state-of-the-art results on these datasets. 
\section{Acknowledgements}
This work was supported in part by a Meta contract to Ohio State University, the Ohio Supercomputer Center, the NCSA Delta Supercomputer Center (OCI 2005572), and the Pittsburgh Supercomputer Center (NSF ACI-1928147).

\bibliographystyle{IEEEtran}

\bibliography{mybib_byyear}


\end{document}